\documentclass[%
 aip,
 amsmath,amssymb,
 reprint,%
]{revtex4-2}

\usepackage{graphicx}
\usepackage{dcolumn}
\usepackage{bm}

\usepackage[utf8]{inputenc}
\usepackage[T1]{fontenc}
\usepackage{mathptmx}
\usepackage{etoolbox}
\usepackage{xcolor}
\usepackage{hyperref}
\usepackage{gensymb}

\usepackage{subcaption}  

\makeatletter
\def\@email#1#2{%
 \endgroup
 \patchcmd{\titleblock@produce}
  {\frontmatter@RRAPformat}
  {\frontmatter@RRAPformat{\produce@RRAP{*#1\href{mailto:#2}{#2}}}\frontmatter@RRAPformat}
  {}{}
}%
\makeatother
\begin{document}

\preprint{AIP/123-QED}

\title{Development and Performance Analysis of Glass-Based Gas-Tight RPCs for Muography Applications}
\author{S. Ikram}
\affiliation{%
Centre for Cosmology, Particle Physics and Phenomenology (CP3), Universit\'e catholique de Louvain, Louvain-la-Neuve, Belgium}
 \email{sumaira.ikram@uclouvain.be}

\author{S. Basnet}
\altaffiliation[Now at ]{Tsung-Dao Lee Institute and School of Physics and Astronomy, Shanghai Jiao Tong University, Shanghai 201210, China}
\affiliation{%
Centre for Cosmology, Particle Physics and Phenomenology (CP3), Universit\'e catholique de Louvain, Louvain-la-Neuve, Belgium}

\author{E. Cortina Gil}
\affiliation{%
Centre for Cosmology, Particle Physics and Phenomenology (CP3), Universit\'e catholique de Louvain, Louvain-la-Neuve, Belgium}

\author{P. Demin}
\affiliation{%
Centre for Cosmology, Particle Physics and Phenomenology (CP3), Universit\'e catholique de Louvain, Louvain-la-Neuve, Belgium}

\author{R.M.I.D. Gamage}
\altaffiliation[Now at ]{Department of Physics, University of Naples 'Federico II', Naples, Italy}
\affiliation{%
Centre for Cosmology, Particle Physics and Phenomenology (CP3), Universit\'e catholique de Louvain, Louvain-la-Neuve, Belgium}

\author{A. Giammanco}
\affiliation{%
Centre for Cosmology, Particle Physics and Phenomenology (CP3), Universit\'e catholique de Louvain, Louvain-la-Neuve, Belgium}

\author{R. Karnam}%
\affiliation{%
National Institute of Science Education and Research Bhubaneswar, An OCC of HBNI, Jatni, Odisha - 752050, India
}%

\author{V. K. S. Kashyap} 
\affiliation{%
National Institute of Science Education and Research Bhubaneswar, An OCC of HBNI, Jatni, Odisha - 752050, India
}%

\author{V. Kumar}
\altaffiliation[Now at ]{Department of Nuclear and Particle Physics, University of Geneva, Switzerland}
\affiliation{%
Centre for Cosmology, Particle Physics and Phenomenology (CP3), Universit\'e catholique de Louvain, Louvain-la-Neuve, Belgium}

\author{B. Mohanty}%
\affiliation{%
National Institute of Science Education and Research Bhubaneswar, An OCC of HBNI, Jatni, Odisha - 752050, India
}%

\author{M. Moussawi}
\altaffiliation[Now at ]{Muon vision, Inc, Cambridge MA, USA}
\affiliation{%
Centre for Cosmology, Particle Physics and Phenomenology (CP3), Universit\'e catholique de Louvain, Louvain-la-Neuve, Belgium}

\author{A. Samalan}
\altaffiliation[Now at ]{Paul Scherrer Institute, Forschungsstrasse 111, 5232 Villigen PSI, Switzerland}
\affiliation{Department of Physics and Astronomy, Ghent University, Ghent, Belgium}

\author{M. Tytgat}
\affiliation{Inter-University Institute for High Energies, Vrĳe Universiteit Brussel, Brussels, Belgium}
\affiliation{Department of Physics and Astronomy, Ghent University, Ghent, Belgium}

\date{\today}

\begin{abstract}

To achieve high-resolution muography of compact targets in scenarios with complex logistical constraints, we are developing a portable muon detector system utilizing glass Resistive Plate Chambers (RPCs). Although RPCs are well understood and widely used, our work focuses on developing a gas-tight variant specifically tailored for a broad range of muography applications, with key design goals including portability, robustness, autonomy, versatility, safety, and cost-effectiveness. Our RPC detectors are designed with various configurations, each featuring unique characteristics and performance attributes. We investigate the temporal evolution of the surface resistivity of glass electrodes, as well as the detector efficiency at varying voltages and thresholds, over a span of several months. These RPCs have been utilized in a small-scale feasibility study on muon absorption using lead blocks.\end{abstract}

\maketitle
\section{Introduction}\label{sec1}

While imaging techniques employing radiation sources such as X-rays and neutrons have demonstrated success in many contexts, their feasibility diminishes when the objects of interest are too large or not transportable. An alternative approach, gaining popularity in recent years, is the use of naturally occurring cosmic-ray muons for imaging large-scale structures through absorption or scattering—a technique known as muography~[\onlinecite{Bonechi_2020}].

The goal of our project is to develop a portable muon telescope customized for applications where the deployment of detectors is constrained by limited space availability, remoteness of location or other logistical issues. Potential use cases include speleology, archaeometry, mineral exploration, and the non-invasive imaging of cultural heritage, such as monumental statues and architectural decorations. 
In this context, our goal includes portability, robustness, safety and autonomy. Portability imposes strong limitations in both weight and size~[\onlinecite{gamage2022}]. To ensure reliable performance across varying ambient conditions, robustness is essential, while safety remains a critical factor, especially in confined spaces. To meet these challenges, we are developing compact and autonomous muon detector prototypes optimized for imaging relatively small volumes of interest in complex and logistically demanding environments. 
Our detectors are based on Resistive Plate Chambers (RPC), chosen for their convenient trade-off between performance on one side and cost and complexity of construction on the other side~[\onlinecite{MuographyBook}], operated in sealed mode to minimize risks associated with the presence of gas sources in confined volumes~[\onlinecite{Basnet_2020}].

This article presents an evaluation of the detector's performance and stability, and is organized as follows: Section~\ref{sec2} describes the setup used in the performance measurements reported in this paper, while Section~\ref{sec3} presents stability and performance studies. Section~\ref{sec4} showcases the findings from a muon absorption measurement in a lead block, an important step to assess the potential of our current prototypes for muography applications. Lastly, Section~\ref{summary} summarizes this study and outlines future directions for the project.

\section{RPC Design and Laboratory Setup}\label{sec2}

This paper reports on three prototypes, indicated as A, B and C, that were built and characterized at the various partner institutes of this project. 
Our detectors are built in different configurations, which enables us to evaluate the impact of subtle variations in design choices. 
These detectors operate using a CMS-RPC gas mixture composed of of 95.2\% Freon, 0.3\% SF6, and 4.5\% isobutane at atmospheric pressure.


For prototype A, chambers are equipped with a charge-sensitive front-end board borrowed from the Muon-RPC system of the CMS experiment~[\onlinecite{Abbrescia:2000sq}]. The board comprises a charge-sensitive Application-Specific Integrated Circuit (ASIC) combined with a Field-Programmable Gate Array (FPGA). For prototype B,the data-acquisition was based on the CAEN DT5550W~[\onlinecite{bworld}] complete DAQ board comprising a PETIROC ASIC and FPGA. The ASIC is responsible for processing charge signals, while the FPGA handles data processing and event triggering. The high-voltage (HV) supply ensures stable operation of the detector components. A detailed description of the CMS readout board and the complete DAQ system is provided in Refs.~[\onlinecite{samalan2023,gamage2022,kumar2024}]. 
Prototype C was mainly tested using a VME-based data-acquisition system, where a CAEN V792 QDC equipped with two A992 for 50~$\Omega$ impedance matching was used to determine the RPC avalanche charges and chamber efficiency.
The prototypes have been tested under various voltage and charge threshold configurations.

\subsection{Prototype A}
\begin{figure}
    \centering
    \includegraphics[width=0.4\textwidth]{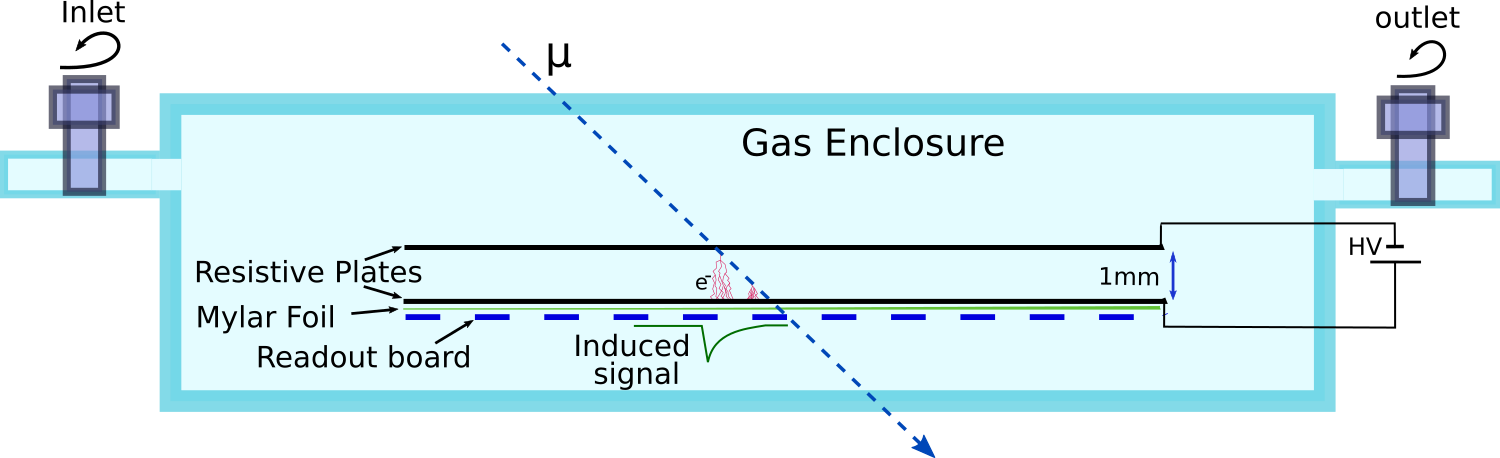}
    \caption{Schematic of glass based RPC placed inside a gas enclosure~[\onlinecite{Vishal_kumar2024}].}
    \label{schematic}
\end{figure}

The RPCs are constructed using $20 \times 20 ~{cm}^2$ glass plates with a thickness of $1.1~mm$. These plates have an active area of  $16 \times 16 ~{cm}^2$, coated with a resistive material consisting of paints with a surface resistivity ranging from 0.5 to 1.0 $M\Omega/\square$. A uniform gas gap of $1~mm$ is maintained between the plates. As shown in Fig.~\ref{schematic}, the plates along with a readout board are placed inside an aluminum gas enclosure containing a standard gas mixture. The enclosure contains a gas volume of 1.9 $L$, with internal dimensions of approximately $32.5\times24.0\times2.45~{cm}^3$. The chamber is flushed 4~--~5 times at a gas flow rate of 0.2 $L/min$ to ensure the thorough removal of residual air. After flushing, the chamber is sealed by closing the inlet and outlet gas valves. The current readout configuration features 16 strips, each measuring 18 $cm$ in length, 0.9 $cm$ in width with a pitch of 1 $cm$~[\onlinecite{Vishal_kumar2024}]. 

\subsection{Prototype B}
\begin{figure}
    \centering
    \includegraphics[width=0.3\textwidth]{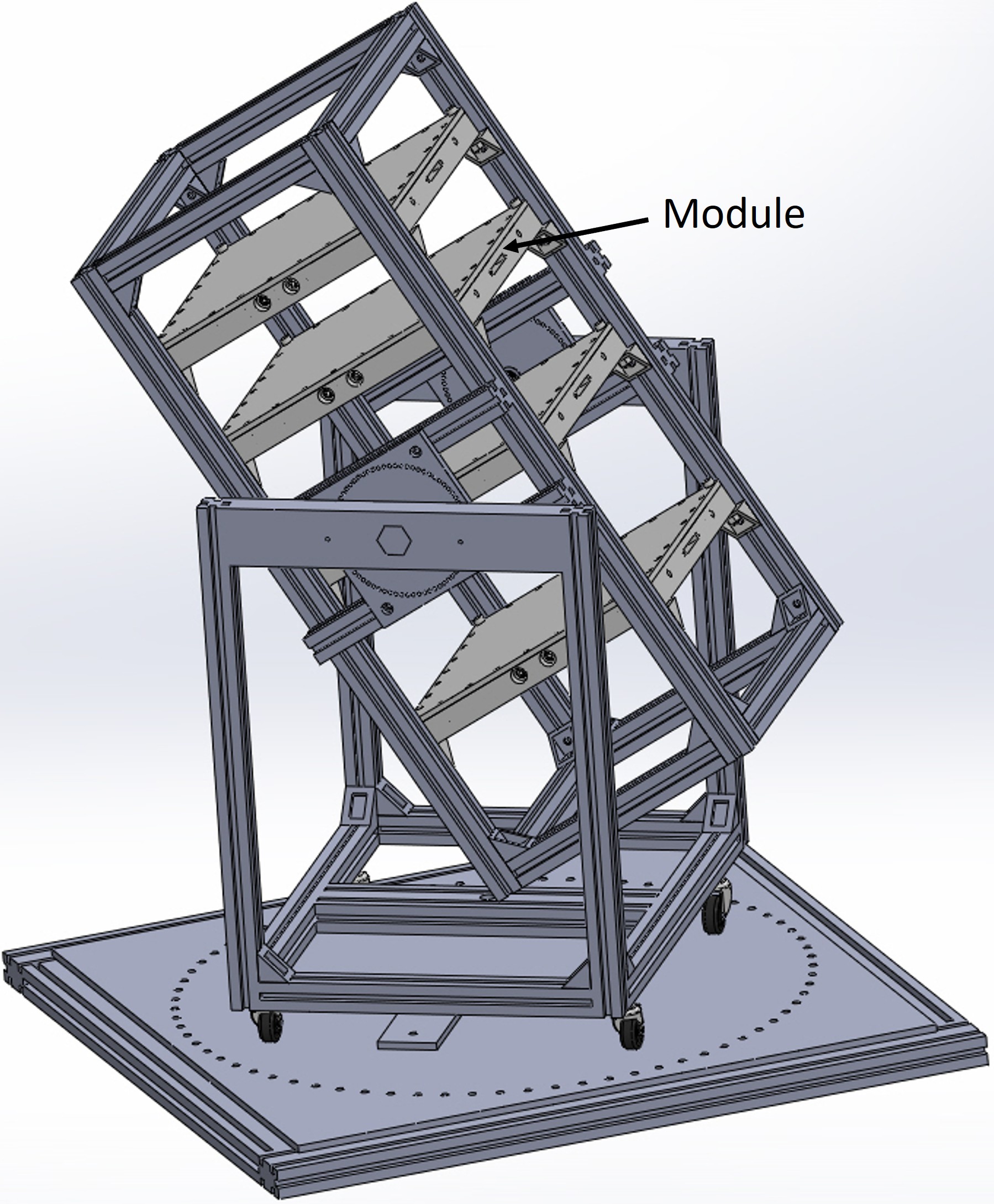}
    \caption{Conceptual design of the muon telescope that is under construction. (Prototype B).}
    \label{fig2a}
\end{figure}

\begin{figure}
    \centering
    \includegraphics[width=0.4\textwidth]{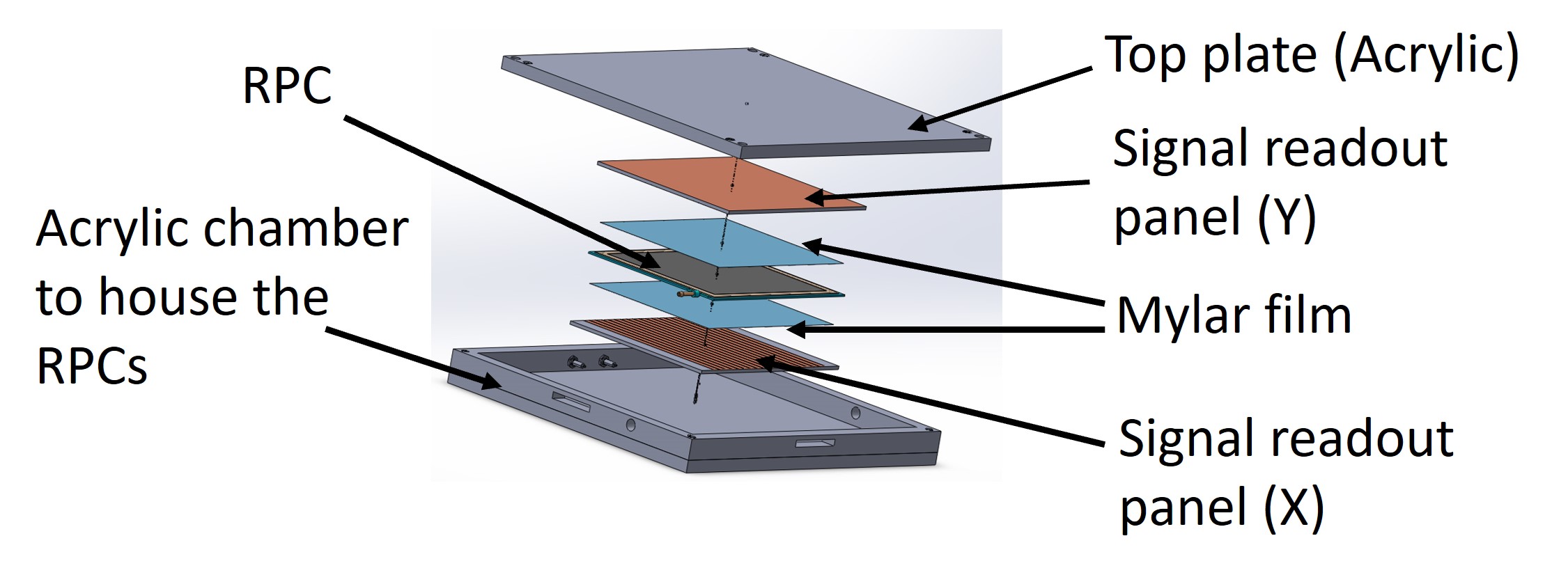}
    \caption{Conceptual design of the RPC module for prototype B.}
    \label{fig2b}
\end{figure}

Prototype B is a muon telescope built in a modular form with four layers as shown in FIG.~\ref{fig2a}. Each layer holds a glass RPC as a tracking detector element that is housed in an airtight casing constructed with acrylic material (Polymethyl methacrylate, PMMA). The RPC is sandwiched between two orthogonally placed Printed Circuit Board (PCB) based strip readout panels, to provide X and Y position of the muon hit as can be seen in Figure~\ref{fig2b}. The RPCs are built using $20 \times 20 \times 0.3$ $cm^3$ size glass electrodes. These are 0.2~cm wide single gas gaps and standalone such that the gas flows only between the electrodes. The readout board has sixteen copper strips with 1~$cm$ pitch on a 0.14~$cm$ thick FR4 sheet.

Further, the four modules are stacked inside an aluminium frame mounted on an alt-azimuthal platform which allows the telescope to rotate in steps of 5$\degree$ with respect to both azimuth and zenith angles. This feature allows the telescope to point in the direction of the volume of target under investigation.   

\subsection{Prototype C}

Following the validation of initial single-gap chambers, a more advanced double-gap glass RPC chamber, prototype C, was constructed. It features a compact design with minimal gas volume   and improved gas tightness plus 2D readout. The chamber contains two 1~$mm$ gas gaps, each made of 1.1~$mm$ thick float glass electrodes with a $28 \times 28$~$cm^2$ active area. The cable-free gas volume is formed using a 3D printed frame that surrounds the glass stack, and that is sandwiched between a top and bottom PCB.
In the current version, each PCB contains 32 8~$mm$ pitch readout strips, with the strips in the top and bottom PCB oriented in orthogonal directions, enabling a 2D XY readout. In addition to the signal traces connecting the strips to the readout connectors outside the gas volume, the PCB also contains traces for the HV and ground connections, and for a pressure \& temperature sensor. 
The glass electrodes were coated with a two-component resistive paint mixture using a manual spray painting technique~[\onlinecite{AmruthaPhD}].
The coatings exhibited surface resistivity values in the range 750-950~$k\Omega$/$\square$ and 1450-1750~$k\Omega$/$\square$ for the top and bottom gas gap respectively.

\begin{figure}
\includegraphics[width=0.4\textwidth]{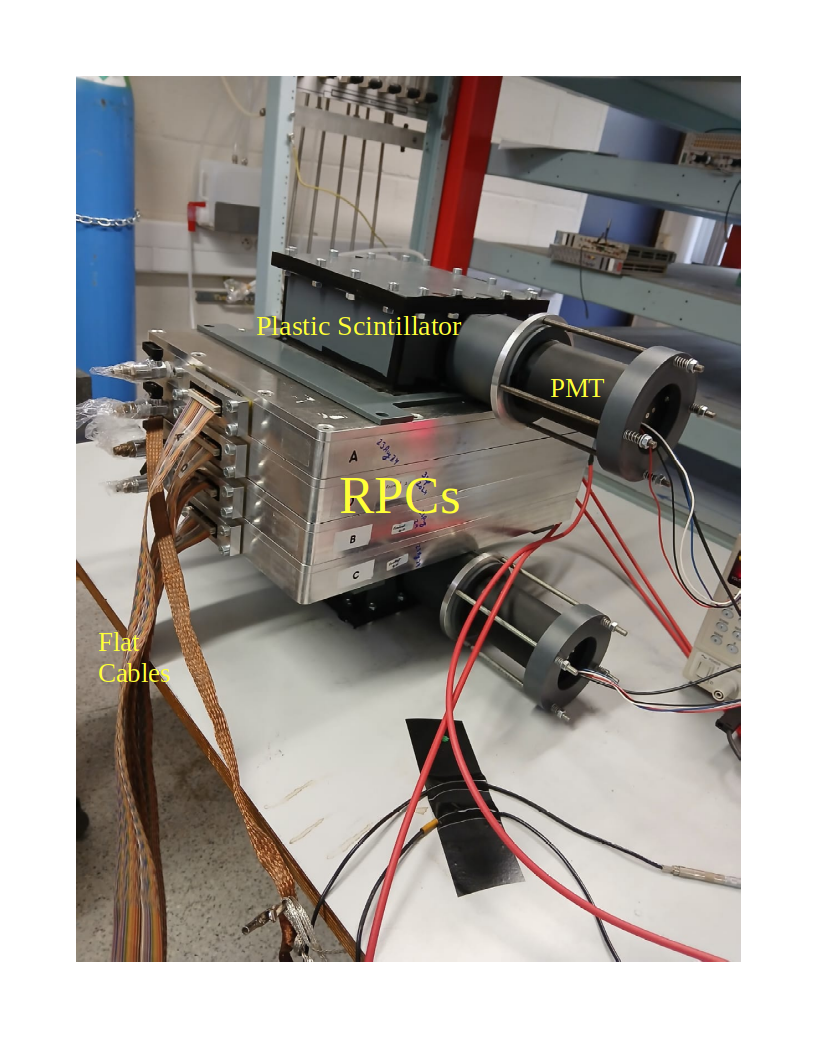}
\caption{Experimental setup for efficiency measurement with prototype A, where RPCs are placed in between two plastic scintillators.}
\label{proto-A-eff-meas-setup}
\end{figure}

\section{Stability and Performance}\label{sec3}

\subsection{Surface Resistivity Monitoring}

 The resistive layers in RPC detectors are critical for receiving and uniformly distributing the operating voltage across the electrode surface, ensuring a homogeneous electric field within the gas gap. They also facilitate charge dissipation during avalanche events~[\onlinecite{MarwaThesis}]. 

\begin{figure}
\includegraphics[width=0.4\textwidth]{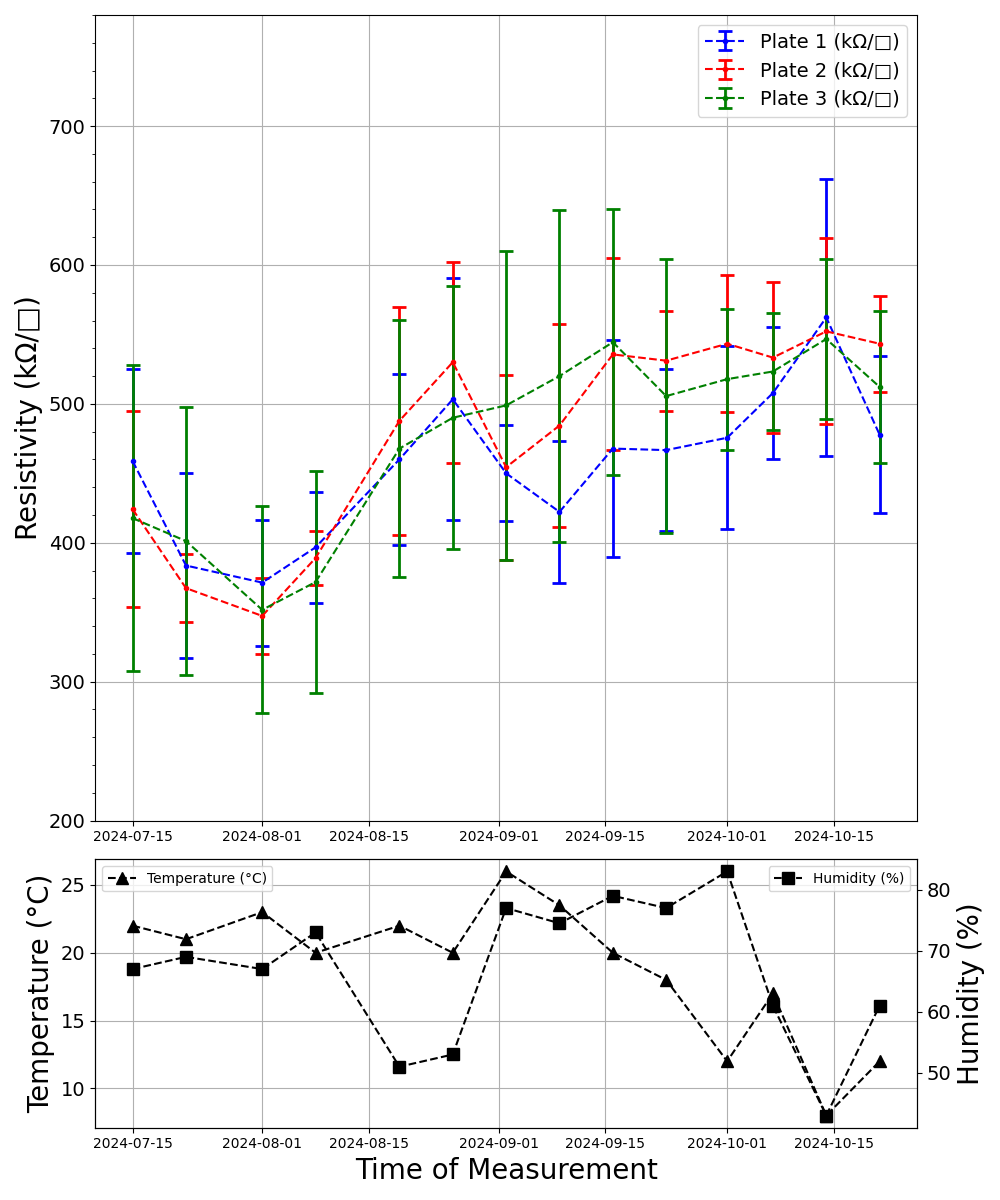}
\caption{Surface resistivity measurements over time for three prototype A plates manufactured using the serigraphy technique (Top). Temperature and humidity recorded during the measurement period (Bottom).}
\label{fig4}
\end{figure}

 The layers currently used for prototype A were fabricated using serigraphy, an industrial coating technique commonly referred to as silk screen printing~[\onlinecite{samalan2023}]. The temporal evolution of the average surface resistivity was studied for three selected glass plates, alongside external measurements of temperature and humidity. Fluctuations in surface resistivity measurements can be attributed to variations in environmental parameters. Figure~\ref{fig4} upper panel shows the surface resistivity measurements over time for three plates manufactured using the serigraphy technique. The bottom panel exhibits variation of temperature and humidity recorded during the measurement period. The results demonstrated stable surface resistivity, highlighting the consistency and reliability of the resistive layers. The surface resistivity measurements are presented with error bars, representing standard deviations arising from location-based variations. Furthermore, the uniformity of the resistive coating, validates the quality and robustness of the fabrication process. 

\begin{figure}
\centering
\includegraphics[width=0.4\textwidth]{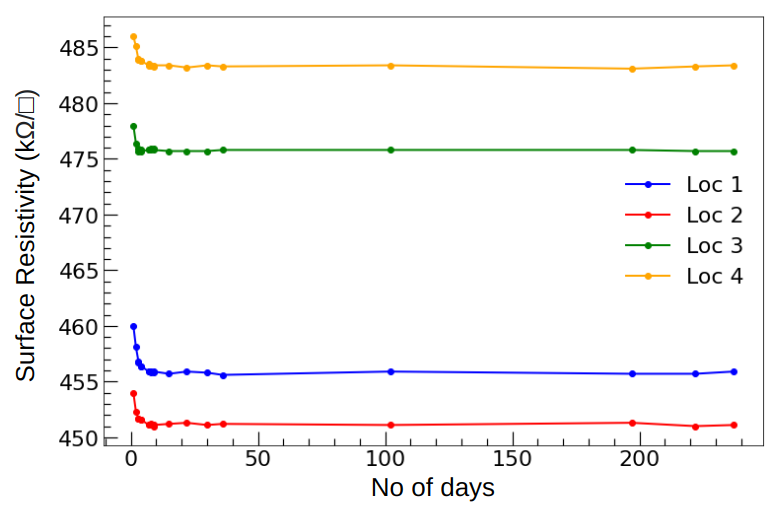}
\caption{Result of a long-term monitoring of surface resistivity values for prototype C, measured at four different locations across the surface of a $30\times30$~$cm^2$ glass plate coated with the spray painting technique.}
\label{fig:protC_resist}
\end{figure}

Figure~\ref{fig:protC_resist} shows results of a 6-months long monitoring of surface resistivity values for a glass plate that was coated with the spray painting technique that is used to construct prototype C. The results indicate the stability and durability of the coating layers obtained with this technique. The measurements at four locations across the glass plate surface also indicate the homogeneity of the resistivity values being around 10\%. Measurements with a Scanning Electron Microscope~[\onlinecite{AmruthaPhD}] showed that the manual spraying technique yields a very uniform coating thickness in the range of 25-30~$\mu m$, with variations within each plate of a few $\mu m$ only.  

\subsection{Performance Studies}

\begin{figure}[!b]
    \centering
    \begin{subfigure}[b]{0.4\textwidth}
        \centering
        \includegraphics[width=\textwidth]{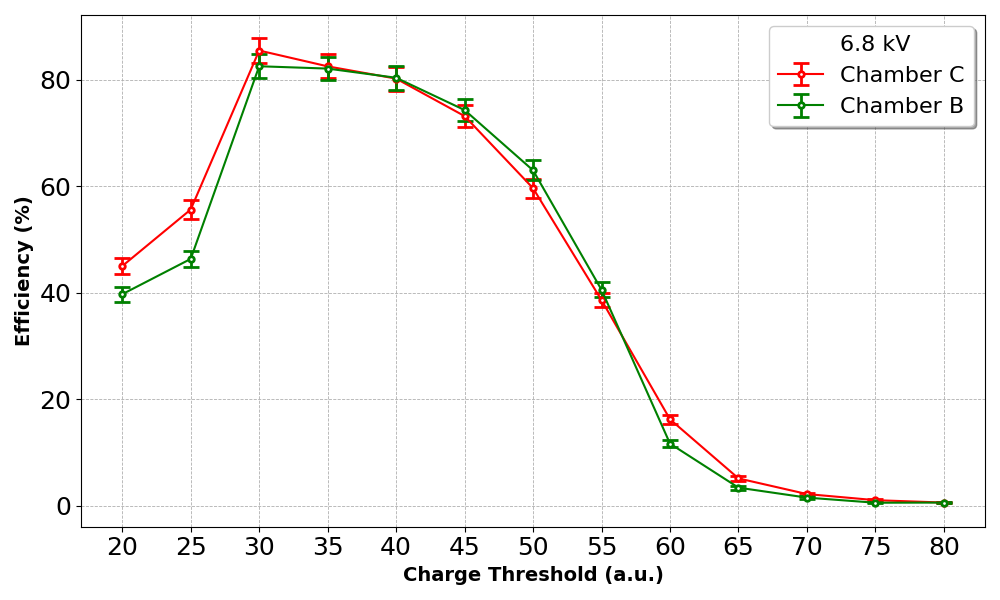} 
        \caption{}
        \label{fig5a}
    \end{subfigure}
    \hfill
    \begin{subfigure}[b]{0.4\textwidth}
        \centering
        \includegraphics[width=\textwidth]{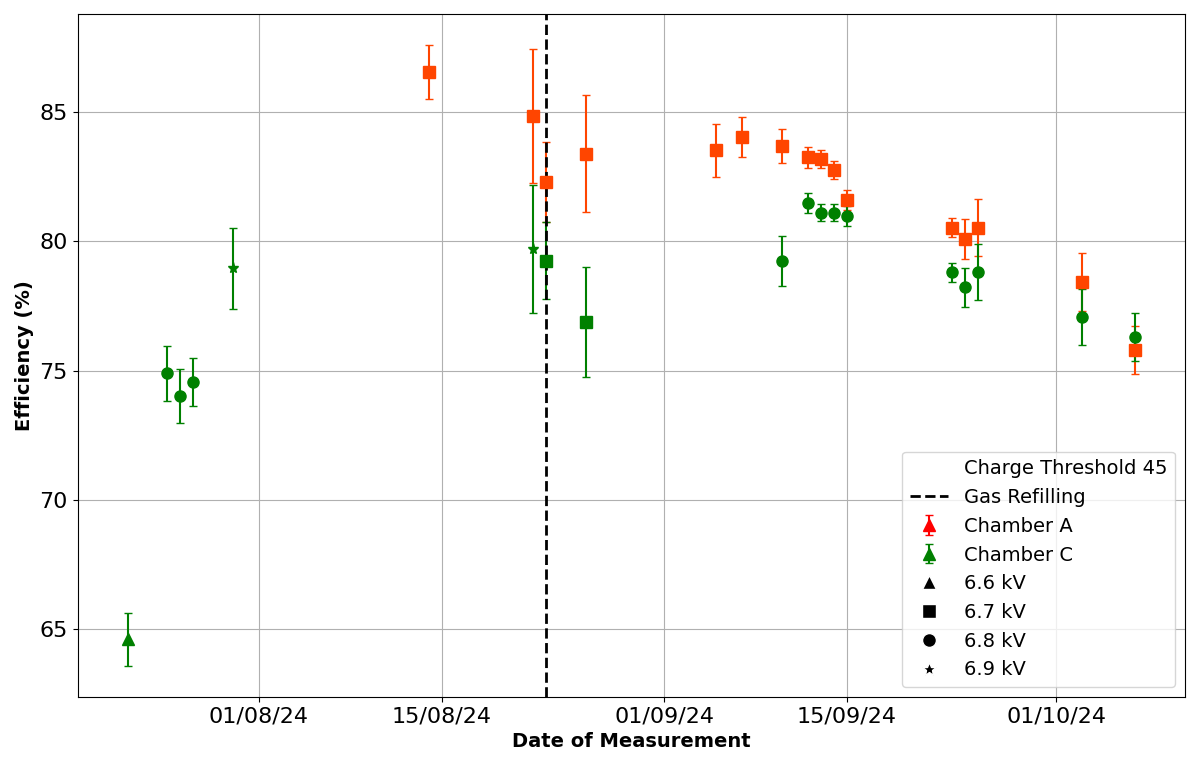}
        \caption{}
        \label{fig5b}
    \end{subfigure}
    \caption{Efficiency variations for two identical prototype A chambers (a) as a function of threshold at HV of 6.8 $kV$ after 3 months of operation; (b) over a span of four months at a fixed threshold of 45 ($a.u$).}
\end{figure}

The detection efficiency of Prototype A was quantified by correlating muon events recorded by the RPCs with those detected in the plastic scintillator, while implementing several noise suppression filters. Figure~\ref{proto-A-eff-meas-setup} shows the experimental setup for efficiency measurement with RPCs placed in between two plastic scintillators. In this configuration, the RPCs are positioned between plastic scintillators, which detect coincident signals to ensure that only authentic muon events are analyzed. A time window filter of $15~ns$ was employed to reject signals outside the intrinsic time resolution of the RPC. A strip multiplicity filter, rejecting events with more than two consecutive strips fired in the same RPC, was applied to constrain the dispersion of the lateral charge. Furthermore, an event filter was applied to isolate single-muon events, eliminating ghost signals associated with strip-based readout systems, as documented in Ref.~[\onlinecite{kumar2024}]. Figure~\ref{fig5a} indicates that the efficiency at a low threshold is higher for 6.8~$kV$.  The maximum efficiency of approximately 85\%  at a threshold value of 30 ($a.u.$) after three months of operation is observed.

 It is observed that efficiency decreases with increasing charge thresholds. For two identical chambers positioned in the same orientation, the efficiency follows a consistent trend, confirming the reproducibility of the results. 
The variation of efficiency with time is shown in Fig.~\ref{fig5b}. For the gas-tight detector, a reduction in efficiency was observed down to 75\% after four months of continuous operation, indicating the potential impact of prolonged usage on detector performance.

\begin{figure}
    \centering
    \includegraphics[width=0.4\textwidth]{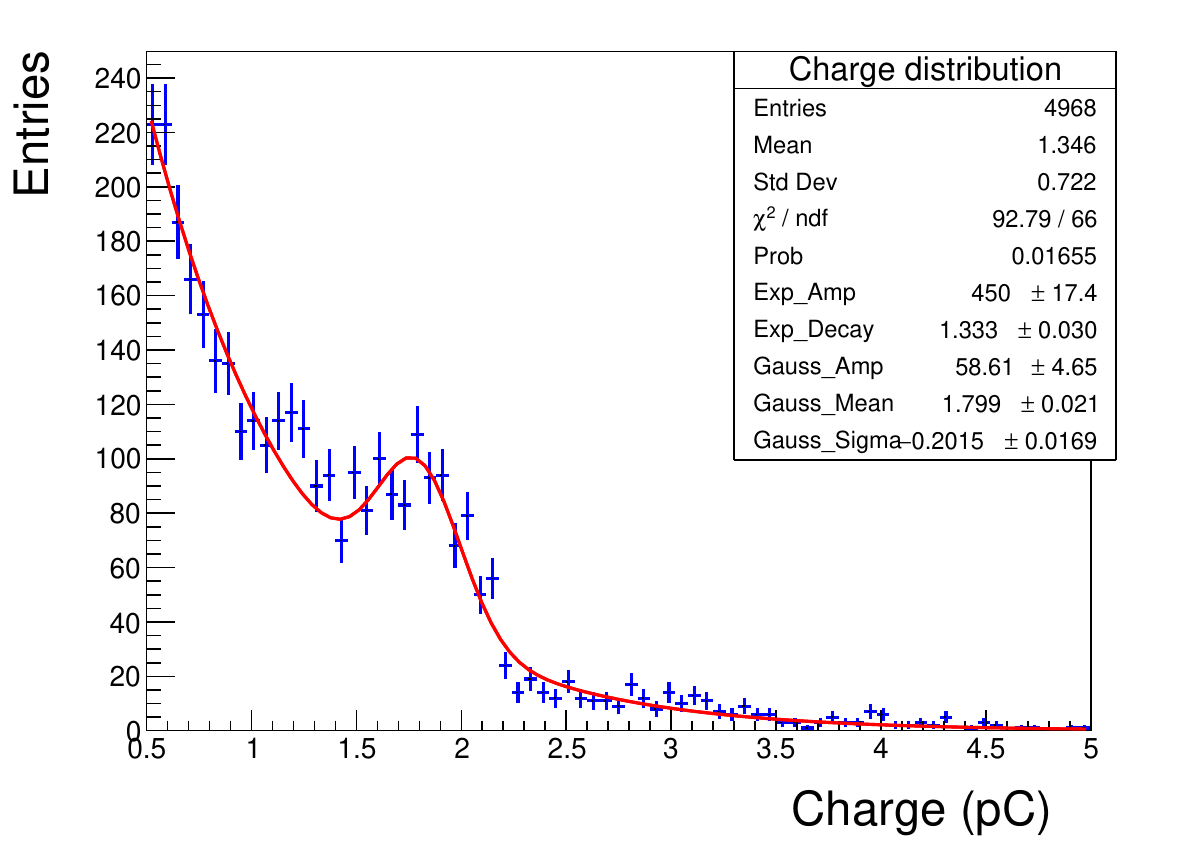}
    \caption{Measured charge collection of the prototype B RPC from a single strip.}
    \label{NISER_RPC_Charge}
\end{figure}

\begin{figure}
    \centering
    \includegraphics[width=0.4\textwidth]{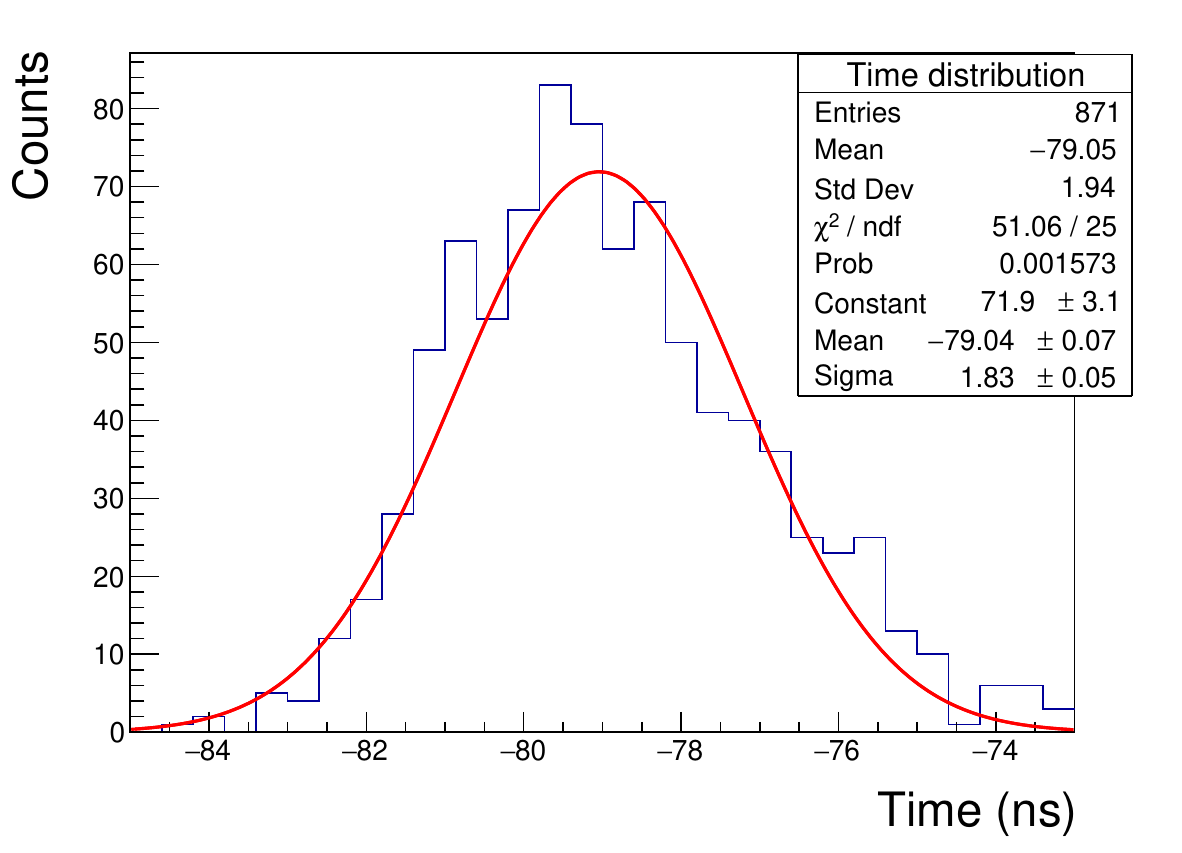}
    \caption{Measured time resolution of the prototype B RPC.}
    \label{NISER_RPC_Time}
\end{figure}

A cosmic-muon coincidence system was built using plastic scintillator detectors to study the performance of prototype B. The cosmic-ray muon detection efficiency of the RPC is measured to be $\geq~$95\%. For more details, please refer to~[\onlinecite{ABHISHEK2025170399}]. Figures~\ref{NISER_RPC_Charge} and~\ref{NISER_RPC_Time} show the measured strip charge collection and time resolution of the RPC are $\sim$ 1.8 $pC$ and $\sim$ 1.8 $ns$ respectively. 

\begin{figure}
    \centering
    \includegraphics[width=0.4\textwidth]{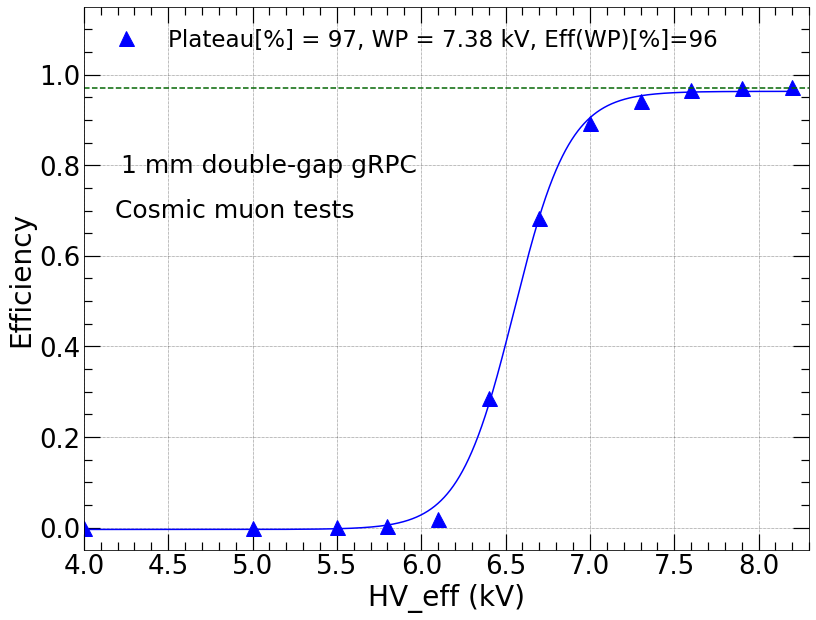}
    \caption{Chamber efficiency for prototype C, measured with cosmic muons as a function of the effective HV.}
    \label{fig:protC_eff}
\end{figure}

Performance studies with prototype C were carried out using a small cosmic-muon coincidence system made of two $16\times 16$~$cm^2$ trigger scintillators~[\onlinecite{AmruthaPhD}]. Due to the small active area of the scintillators, only 18 out of the 32 strips per PCB of the prototype chamber were covered in the measurements.
Results on the cluster multiplicity and RPC cluster charge as function of applied HV have been reported in Ref.~[\onlinecite{samalan2023}]. Fig.~\ref{fig:protC_eff} shows the measured efficiency for prototype C as a function of the effective HV. The efficiency reaches a clear plateau at about 97\% for voltages above 7.5~$kV$, demonstrating the excellent performance of the double-gap chamber layout.

\section{Absorption Muography Using Lead Block}\label{sec4}

\begin{figure}[!b]
\includegraphics[width=0.4\textwidth]{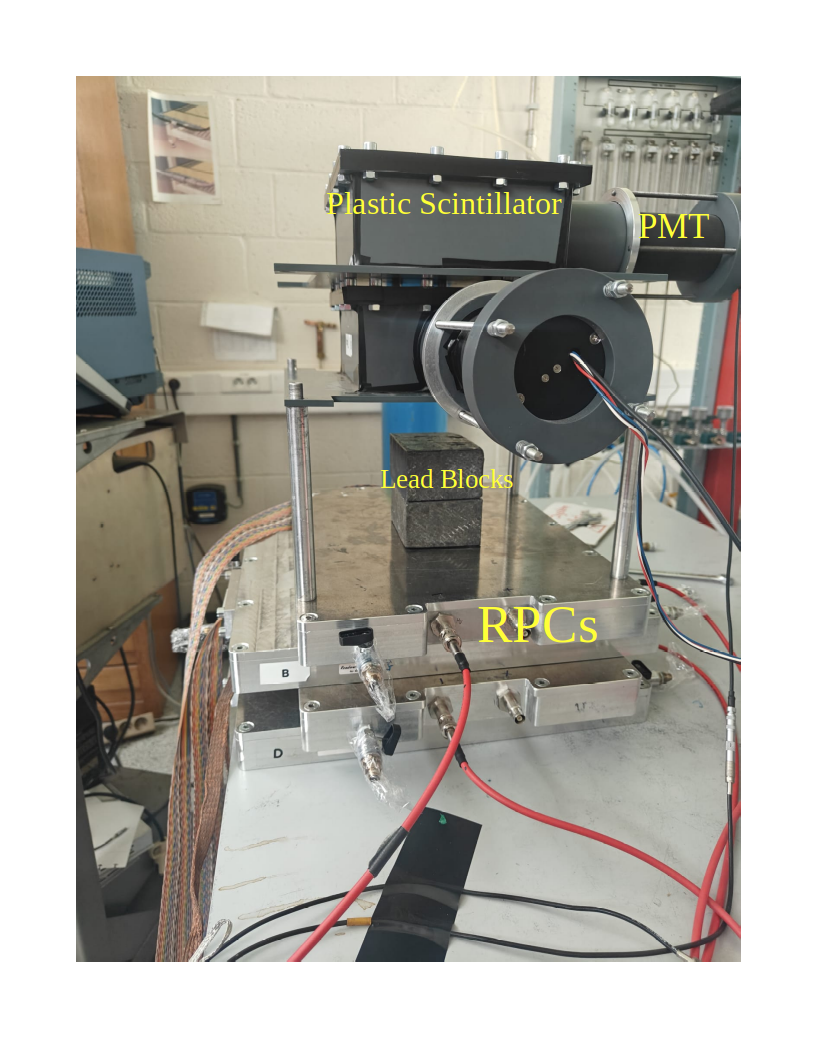}
\caption{Experimental setup for muon absorption tomography with prototype A (Setup 1).}
\label{fig6a}
\end{figure}

\begin{figure}
\includegraphics[width=0.4\textwidth]{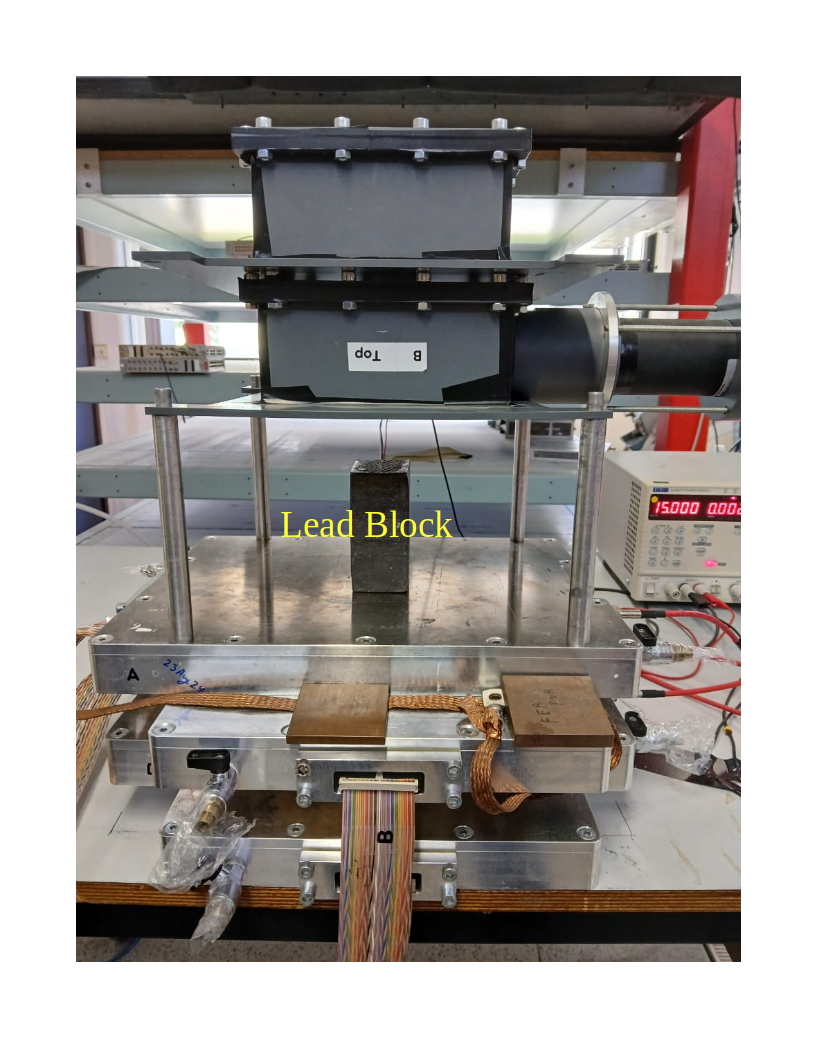}
\caption{Experimental setup for muon absorption tomography with prototype A (Setup 2).}
\label{fig6b}
\end{figure} 

A small-scale feasibility study on muon absorption was performed in the laboratory, using prototype A. The experimental setup included two plastic scintillators positioned at the top and RPCs are placed at the bottom as illustrated in Figs.~\ref{fig6a} and ~\ref{fig6b}. We performed absorption muography using two different configurations. In the Setup 1, absorption in two lead blocks, each measuring $10 \times 7.5\times 4.5 ~{cm}~^3$ were studied. These blocks were placed between the scintillators and RPCs in a stacked configuration. In Setup 2, a single lead block measuring $4.5 \times 7.5 \times 10 ~{cm}~^3$  was positioned vertically. These arrangements allowed us to study muon absorption within the lead blocks under controlled experimental conditions. To evaluate the performance of detectors, along with the DAQ and analysis methodologies, a simplified muon absorption experiment has been conducted. In Figs.~\ref{fig6a} and~\ref{fig6b} two plastic scintillators in conjunction with PMTs positioned $15~cm$ above the RPC chambers supported by four cylindrical aluminum pillars are shown. These pillars serve the dual purpose of offering structural support and defining a scanning volume, commonly referred to as the Region Of Interest (ROI).
The signals obtained from the PMTs serve as trigger signals for data acquisition of the RPCs, similar to the process used in efficiency measurements. For Setup 1, two distinct data sets were acquired: (1) a configuration with two lead blocks positioned within the ROI, and (2) a free-sky measurement (no lead blocks) serving as the baseline reference. An analogous approach was followed for Setup 2, utilizing a single lead block instead. In both experimental setups, two orthogonally arranged RPCs (labeled "Layer 0" and "Layer 1") acquired data for bidimensional (XY) spatial information.

\begin{figure}[h]
    \centering
    \begin{subfigure}[b]{0.4\textwidth}
        \centering
        \includegraphics[width=\textwidth]{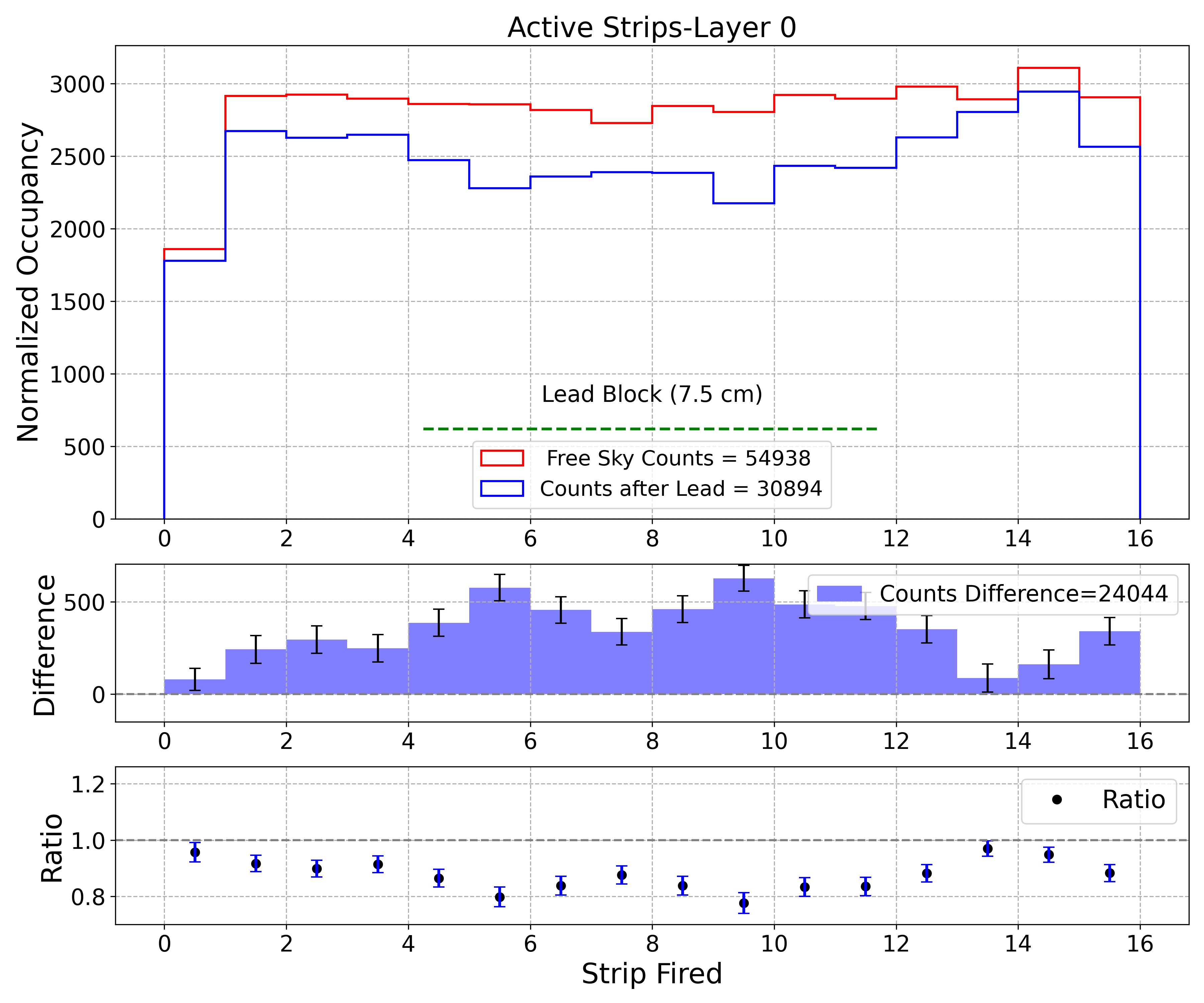}
        \caption{}
        \label{fig7a}
    \end{subfigure}
    \hfill
    \begin{subfigure}[b]{0.4\textwidth}
        \centering
        \includegraphics[width=\textwidth]{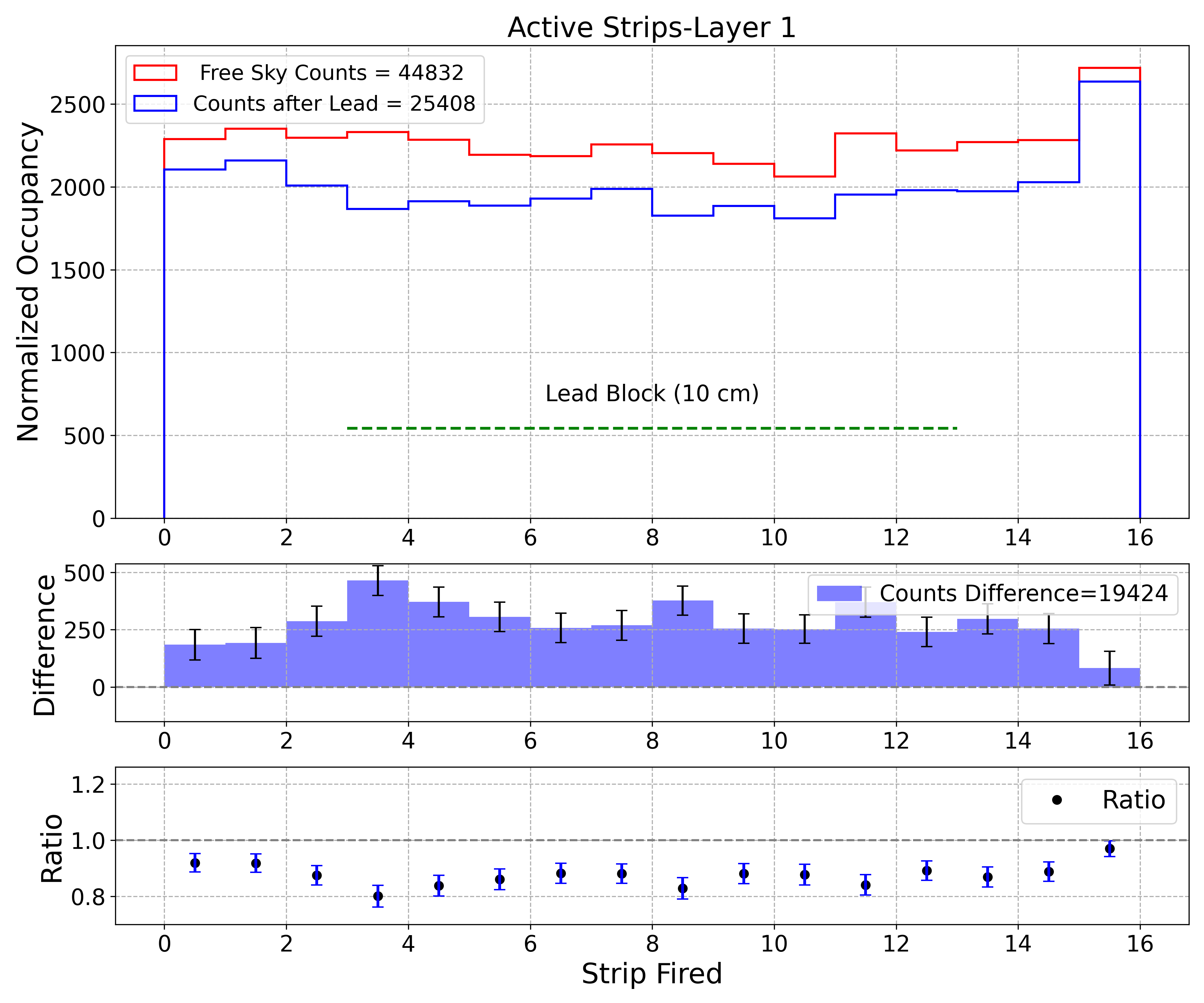}
        \caption{}
        \label{fig7b}
    \end{subfigure}
    \caption{Comparison of active strip in Setup 1 (2 lead blocks), with and without lead blocks in the Region of Interest (ROI). The top panels show the strip occupancy, the middle panels illustrate the difference and the bottom panels present the ratio of normalized occupancy values. Results are displayed for RPC Layer 0 (a) and RPC Layer 1 (b) of prototype A.}
\end{figure}

\begin{figure}[h]
    \centering
    \begin{subfigure}[b]{0.4\textwidth}
        \centering
        \includegraphics[width=\textwidth]{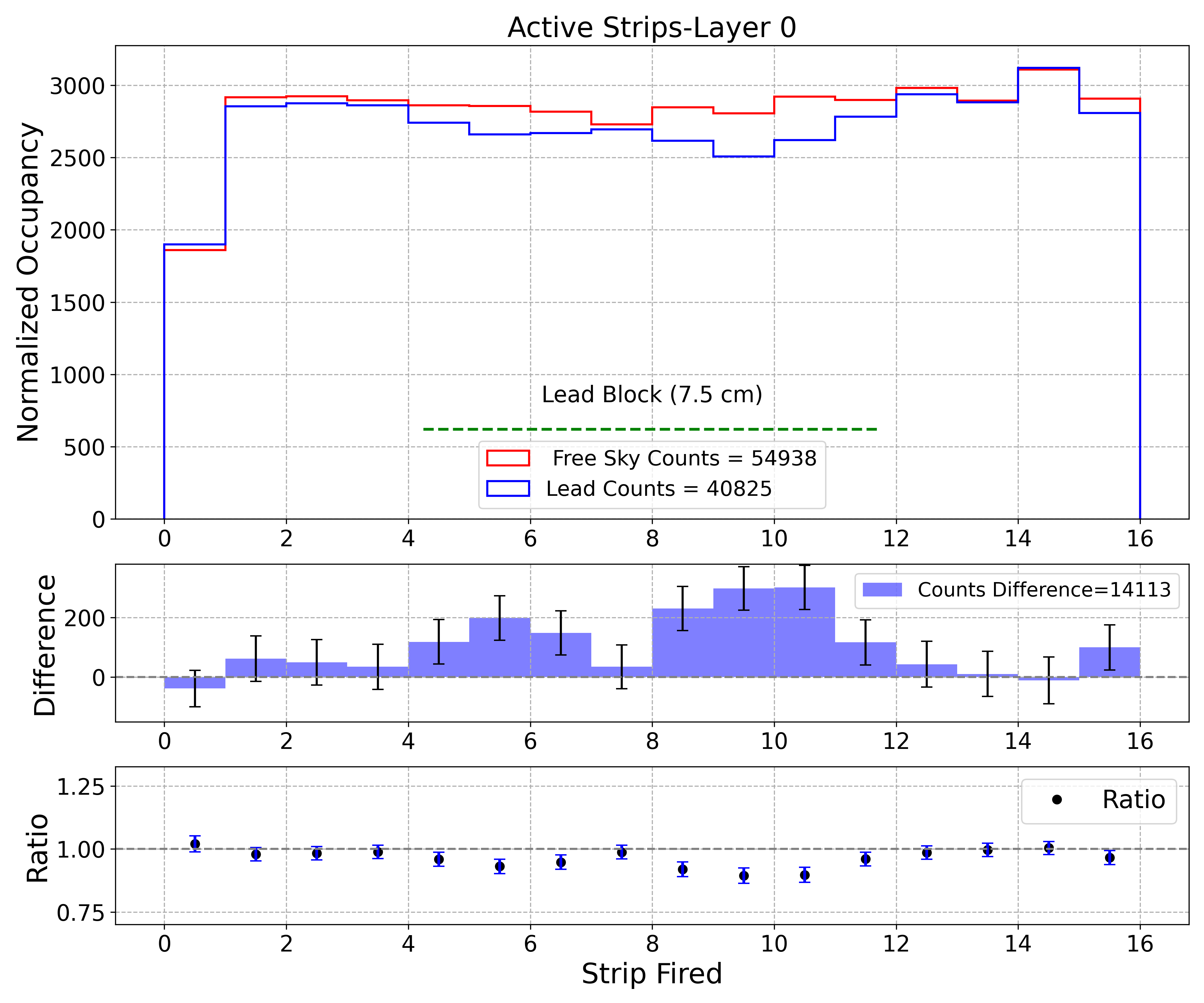}
        \caption{}
        \label{fig8a}
    \end{subfigure}
    \hfill
    \begin{subfigure}[b]{0.4\textwidth}
        \centering
        \includegraphics[width=\textwidth]{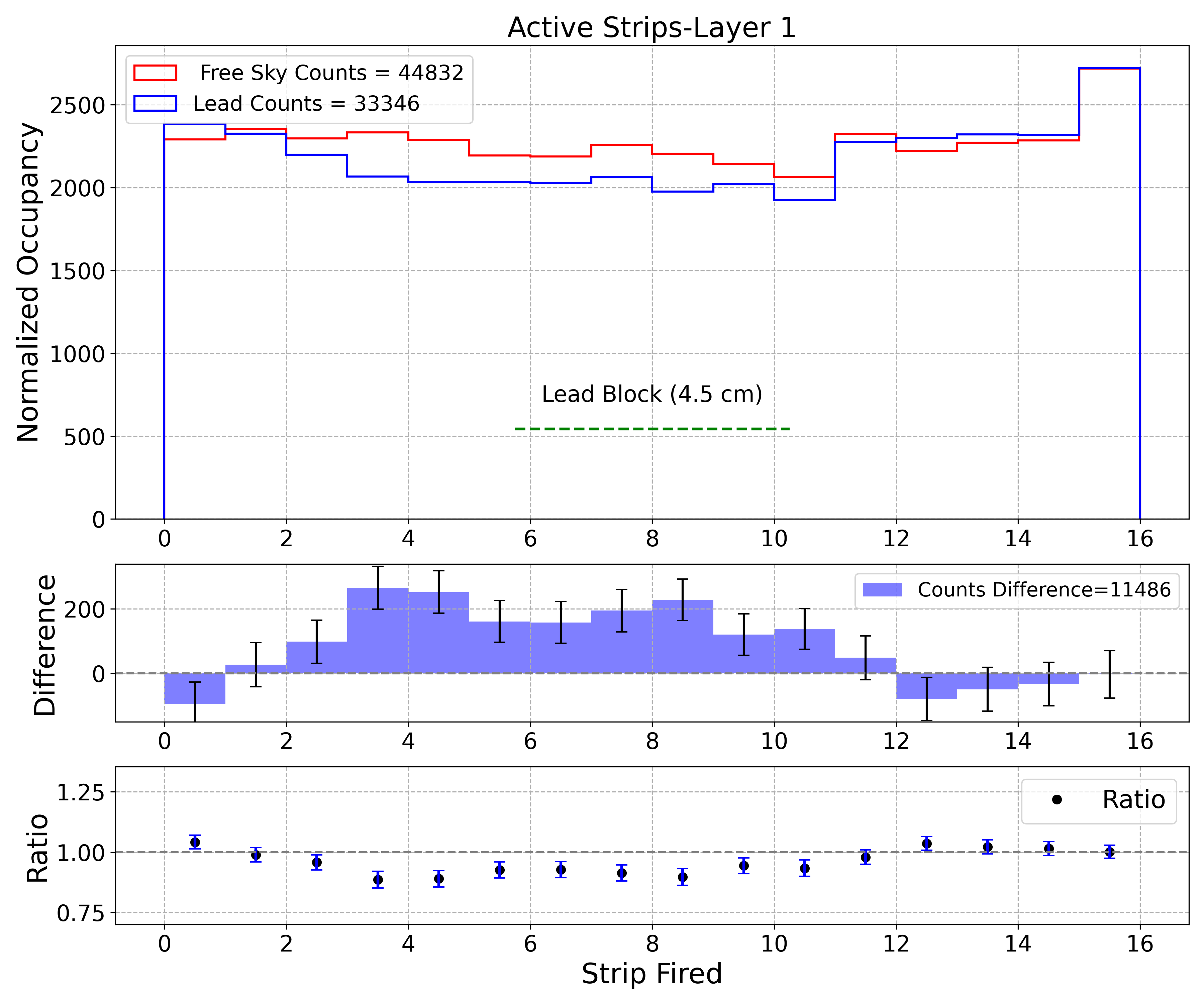}
        \caption{}
        \label{fig8b}
    \end{subfigure}
    \caption{Comparison of active strip in Setup 2 (1 lead block), with and without lead block in the Region of Interest (ROI). The top panels show the strip occupancy, the middle panels illustrate the difference and the bottom panels present the ratio of normalized occupancy values. Results are displayed for RPC Layer 0 (a) and RPC Layer 1 (b) of prototype A.}
\end{figure}

 The normalized occupancy for two scenarios: one with free sky data and the other with the lead block(s) is shown in Figs.~\ref{fig7a}, \ref{fig7b}, \ref{fig8a} and~\ref{fig8b}. Occupancy is normalized by the number of hours of data collection. This normalization is essential for mitigating statistical fluctuations in the muon flux over the data taking period. Examining the normalized occupancy plot reveals a noticeable reduction of muon counts in correspondence of the lead block position and its surroundings.
The attenuation of muon flux due to absorption is analyzed by examining the difference and thratio of normalized strip occupancy. The difference plot illustrates the variation in muon counts across strips before and after placing the lead block, highlighting regions with significant absorption. Similarly, the ratio plot presents the strip-wise proportion of normalized occupancy before and after the placement of lead block(s). These figures clearly demonstrate a reduction in strip occupancy at the position of lead block(s), with additional decrease observed in surrounding regions. This suggests that muons approaching at an angle are either absorbed or scattered away. The green line in the figures marks the exact location of the lead block along the axis perpendicular to the strips.

\section{Summary and Outlook}
\label{summary}

A comprehensive performance evaluation of our gas-tight RPCs has been successfully carried out, utilizing a meticulous pulse processing and data acquisition approach. Additionally, a detailed threshold scan has been conducted across various voltage settings to determine the optimal operating conditions. To further assess the feasibility and practical performance of these gas-tight RPCs, a simplified muon absorption experiment using a two RPCs was performed, yielding valuable insights. Our prototype RPC detectors have undergone long-term stability tests to assess gas performance, a crucial factor influencing overall detector efficiency. These tests have been conducted to validate field performance.

For a next version of prototype A, we plan to further optimize our 3D-printed rigid frame, made from a lightweight material such as carbon fiber or plastic. This frame securely holds the glass plates while maintaining a uniform gap. Additionally, it replaces the aluminum casing, incorporating gas inlet and outlet channels designed to ensure even gas distribution. Compared to aluminum, the lighter frame enhances portability, making the setup more efficient and easier to handle~[\onlinecite{gamage2022}]. 

For the readout of the prototypes A and C, we are transitioning to the MAROC chip, significantly enhancing our system's readout granularities. The MAROC chip is considerably more compact than the CMS chip (O(1)~$cm$ vs. O(10)~$cm$). It offers a substantial improvement in channel capacity, supporting 64 electronic channels per chip — eight times the capacity of the CMS chip, which is limited to just 8 channels. This upgrade will greatly enhance the scalability and performance of our setup~[\onlinecite{Basnet_2020}].
\\

\begin{acknowledgments}
This work was partially supported by the EU Horizon 2020 Research and Innovation Programme under the Marie Sklodowska-Curie Grant Agreement No. 822185, by the Fonds de la Recherche Scientifique
- FNRS (Belgium) under Grants No. T.0099.19, J.0070.21 and J.0043.22, and by the Department of Atomic Energy, India.  B. Mohanty acknowledges the J C Bose Fellowship 
of the Department of Science and Technology, India, for the support.
\end{acknowledgments}

\section*{Conflict of Interest}
The authors have no conflicts to disclose.

\section*{Data Availability Statement}

The data that support the findings of this study are available from the corresponding author upon reasonable request.

\section*{References}
\nocite{*}
\bibliography{aipsamp}

\end{document}